\documentclass[aps,prl,showpacs,10pt,twocolumn,superscriptaddress]{revtex4-2}
\usepackage[unicode,breaklinks]{hyperref}
\usepackage{physics,amsmath,amssymb}
\usepackage{dcolumn}
\usepackage{scalerel}
\usepackage{bm}
\usepackage{graphicx}
\usepackage{tikz}
\usetikzlibrary{shapes}
\usepackage{soul}

\begin{document}

\title{Dark Soliton Formation as a Dark-State Phase Transition in a Dissipative Superfluid Josephson Junction Chain}
\author{Robbe Ceulemans}
\email{robbe.ceulemans@uantwerpen.be}
\affiliation{TQC, Departement Fysica, Universiteit Antwerpen, B-2000 Antwerpen, Belgium}
\author{Samuel E. Begg}
\email{samuel.begg@utdallas.edu}
\affiliation{Department of Physics, Oklahoma State University, Stillwater, Oklahoma 74078, USA}
\affiliation{Asia Pacific Center for Theoretical Physics, Pohang 37673, Korea}
\affiliation{Department of Physics, The University of Texas at Dallas, Richardson, Texas 75080, USA}
\author{Matthew J. Davis}
\email{mdavis@uq.edu.au}
\affiliation{Australian Research Council Centre of Excellence in Future Low-Energy Electronics Technologies, School of Mathematics and Physics, University of Queensland, St Lucia, Queensland 4072, Australia.}
\author{Michiel Wouters}
\email{michiel.wouters@uantwerpen.be}
\affiliation{TQC, Departement Fysica, Universiteit Antwerpen, B-2000 Antwerpen, Belgium}
\date{\today}

\begin{abstract}
We identify and characterize a first-order dark-state phase transition between a discrete dark soliton and a uniform superfluid in a Bose-Hubbard chain with a single lossy site.
Using classical-field (truncated-Wigner) simulations together with a Bogoliubov stability analysis, we show that the dark-state nature of the soliton suppresses fluctuations and shifts the critical point relative to the comparable phenomenon of optical bistability in driven-dissipative Kerr resonators. 
We then demonstrate that this mechanism quantitatively captures the bistability phase boundary observed in the  experiment of R. Labouvie \textit{et al.} [Phys. Rev. Lett. \textbf{116}, 235302 (2016)], resolving substantial discrepancies in prior modeling efforts.
Our results reveal how driving, dissipation and quantum coherence can interact to induce nonequilibrium phase transitions in ultra-cold atomic gases.
\end{abstract}
\maketitle

\textit{Introduction.---}  
While dissipation is often seen as a nuisance for quantum systems, 
engineered dissipation can stabilize coherent many-body states and even drive new nonequilibrium phenomena \cite{Diehl2008,Verstraete2009,Barreiro2011,Brazhnyi2009}. 
Prominent experimental examples include exciton-polariton condensation \cite{Kasprzak2006}, a Mott insulator of photons \cite{Ma2019}, and the Mott-superfluid transition via engineered two-body loss in cold atoms \cite{Tomita2017}. This provides a new frontier to investigate the role of quantum fluctuations in driven-dissipative phase transitions \cite{Sieberer2016,Sieberer2025}, as illustrated by recent work showing that dark states \cite{Muller2012,Harrington2022} can induce phenomena such as absorbing state phase transitions \cite{Marcuzzi2015,Buchhold2017,Roscher2018} and nonequilibrium quantum critical dynamics \cite{Marino2016prl,Marino2016prb,Begg_spin_2024}. A canonical, first-order phase transition is given by the so-called `optical bistability' in driven-dissipative Kerr resonators 
\cite{Drummond1980b}, where switches between macrostates can be induced by quantum fluctuations \cite{rodriguez2017probing}. Analogous transitions have been studied in a wide range of systems \cite{kessler_dissipative_2012,huybrechts_dynamical_2020,roberts_competition_2023,foss2017emergent,Labouvie2016}.\par

\begin{figure}[t]
    \begin{minipage}[b]{0.47\columnwidth}
        \includegraphics[width = \linewidth]{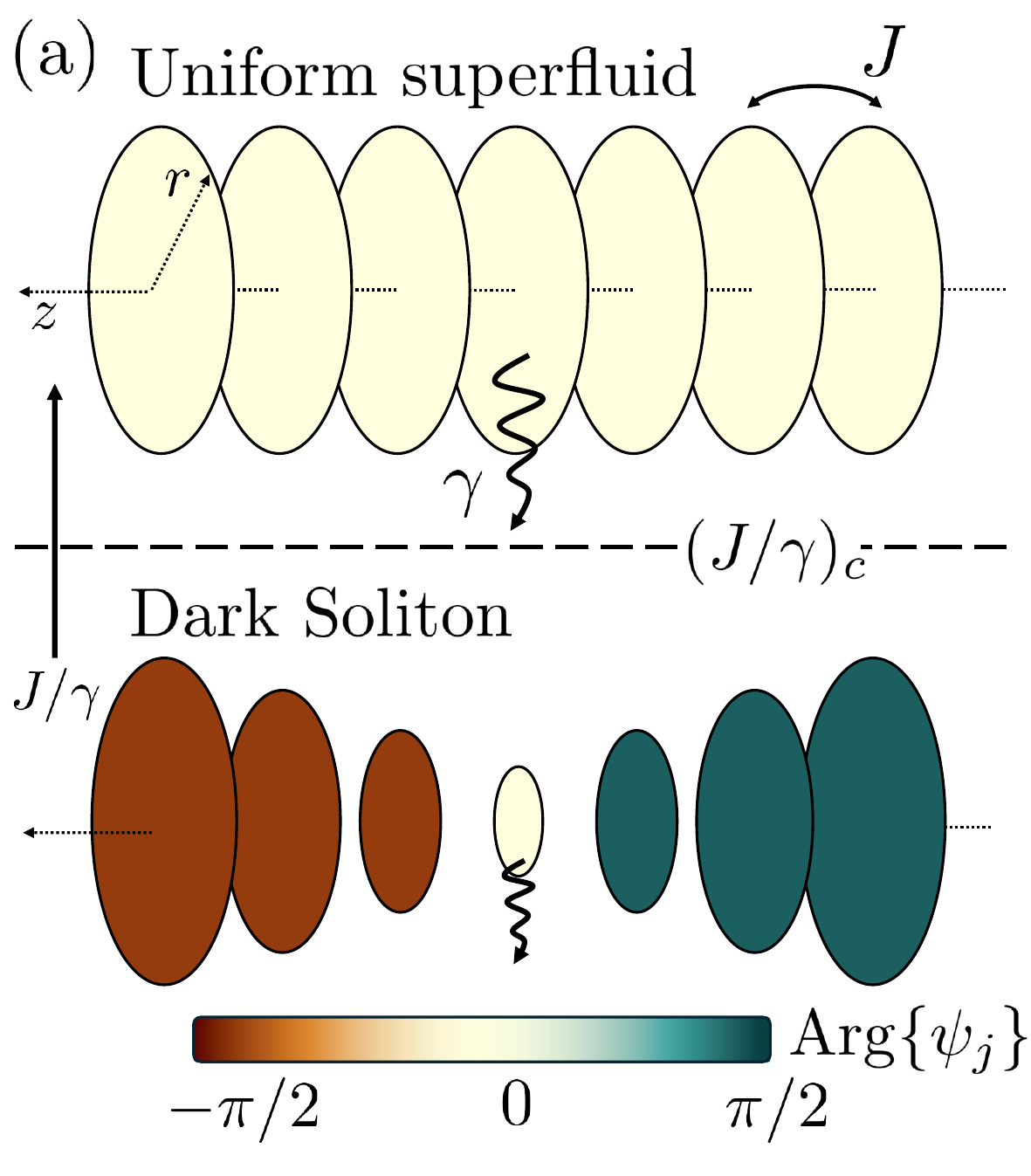}
    \end{minipage}
    \begin{minipage}[b]{0.46\columnwidth} 
        \includegraphics[width = \linewidth]{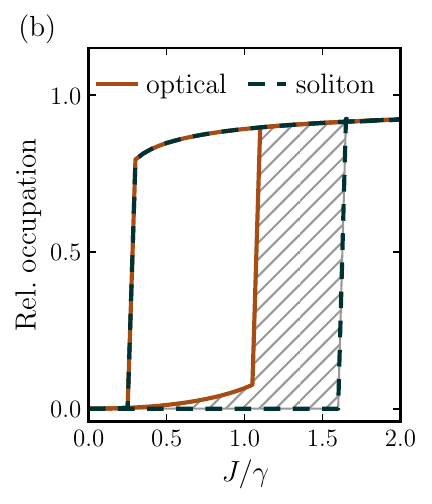}
    \end{minipage} 
    \caption{(a) Illustration of the two late-time macrostates in a quasi-1D Josephson junction chain with tunneling $J$ and local particle loss $\gamma$ on the central site. 
    As $J/\gamma$ increases there is a phase transition from a \textit{dark soliton} (bottom) to a uniform superfluid (top). The former is characterized by a $\pi$ phase jump and vanishing density on the central site, and is a dark state that decouples from the dissipation.
    (b) Central-site occupation $\abs{\psi_0}^2$ (normalized by maximum occupation $N_f$) vs $J$, showing the bistability of numerical mean-field solutions. Dashed lines (green) show the case of a superfluid to dark soliton, which sweeps a bigger area than a comparable transition connecting the superfluid to a symmetric low-density state without $\pi$ phase jump (orange).}\label{fig:setup}
\end{figure}

In this work we demonstrate a first-order dark-state phase transition between a dark soliton and a uniform superfluid in a dissipative Bose-Hubbard (BH) model [Fig.~\ref{fig:setup}(a)]. A dark soliton is an excited eigenmode of the Gross-Pitaevskii equation characterized by a $\pi$ phase slip and vanishing density at its core, which allows the state to decouple from locally engineered loss. 
The transition resembles optical bistability, yet the soliton’s spatial phase texture and dark-state character suppress fluctuations and shift the critical point.
Using semi-classical field models, we further show that this phase transition quantitatively explains recent cold-atom experiments \cite{Labouvie2016,benary_experimental_2022,Rohrle2024} that so far have eluded understanding from first principles. Beyond the implications for experiments, our work identifies a unique nonequilibrium critical phenomenon driven by the interplay of quantum coherence and dissipation. This also complements a growing body of work exploring local loss as a tool for controlling quantum transport \cite{Barontini2013,Mullers2018,Lebrat2019,Froml2019,Kunimi2019,Sels2020,Froml2020,Yanay2020,Muller2021, Will2023,Reeves2021,Huang2023,Visuri2023,Visuri2023b,Gievers2024, Stefanini2025,Despres2025,Ogino2025}, or for preparing far-from-equilibrium superfluid states with exotic relaxation \cite{Labouvie2015,Olsen2016,Fischer2017,  Mink2022,begg_nonequilibrium_2024}.

\textit{Dark-state phase transition.---}
We consider a 1D BH chain of integer length $L+1$ subject to Markovian dissipation \cite{Lindblad1976} in the form of single-particle loss on the central site ($j = 0$) with rate $\gamma$, with $j \in \{-L/2,L/2\}$ the lattice index. In the semiclassical limit the field amplitude $\psi_j$ for $j\neq 0$ obeys \cite{Ceulemans2023}
\begin{equation}
    i\hbar\partial_t\psi_j = \mathcal{H}\psi_j - J\left(\psi_{j-1} + \psi_{j+1}\right) \label{eq:GPEOther},
\end{equation}
while for $j= 0$ the dissipative stochastic evolution is
\begin{equation}
    i\hbar\partial_t\psi_0 = (\mathcal{H}-i\gamma/2)\psi_0 - J\left(\psi_{-1}+\psi_1\right) + \sqrt{\gamma/2}~\eta, \label{eq:GPECentral}
\end{equation}
with tunneling amplitude $J$ and where $\mathcal{H}\psi_j=U\abs{\psi_j}^2\psi_j$ describes the on-site interactions with an effective strength $U$. The complex Gaussian noise $\eta$ satisfies $\langle \eta(t) \rangle = 0$ and  $\langle \eta(t)\eta^*(t^\prime)\rangle =\delta(t-t^\prime)$.
Under the influence of driving from a large number of highly occupied sites on either side, the lossy site swiftly relaxes to a (quasi-)steady state determined by the balance between the loss rate and tunneling amplitude. For weak tunneling, $J/\gamma \ll 1$, losses dominate and the central site settles in a “normal” (quantum Zeno) state with low occupation and condensate fraction. In contrast, at large $J/\gamma$ a superfluid state emerges with the lossy site fully occupied. \par
In the thermodynamic limit $N_j \propto N \rightarrow \infty$, where $N_j$ is the atom number for each site, there is a first-order transition between these `empty' and `full' states at a critical value $(J/\gamma)_c$, for a fixed $\gamma/\mu$ \cite{Ceulemans2023}.
Consequently, the system displays bistability as shown in Fig.~\ref{fig:setup}(b) by the associated hysteresis loop in the steady-state occupation. 
Similar results for a comparable system of coupled superfluid Josephson junctions have been demonstrated experimentally in Refs.~\cite{Labouvie2016,benary_experimental_2022,Rohrle2024}. 
Most previous analyses interpreted the low-density branch as a Kerr-like low-amplitude state [orange curve in Fig. \ref{fig:setup}(b)] with a single pump \cite{Labouvie2016,Sels2020,benary_experimental_2022,Rohrle2024,Ceulemans2023}. Reference \cite{Reeves2021} allowed for two pumps that could dephase, but did not consider their phase coherence and was not able to reproduce the experimental boundaries without changing experimental parameters.
Here, we provide strong evidence that the local `empty' state actually corresponds to a spatially extended dark soliton.

The discrete dark soliton has the characteristic spatial profile \cite{kivshar_dark_1994}
\begin{equation}
    \psi_j = \sqrt{N_f}\tanh\left[jd_z/\sqrt{2\xi^2}\right],
    \label{eq:soliton}
\end{equation}
where $d_z$ is the lattice spacing, $N_f$ is the population of a full site, and $\xi = \sqrt{Jd_z^2/\mu}$ is the soliton’s healing length. It exhibits a $\pi$ phase jump and vanishing density at the central site [Fig. 1(a)]. Together with the fact it is an eigenmode of the discrete 1D Gross-Pitaevskii equation, this ensures that the state completely decouples from the bath, making the dark soliton a dark state of Eqs.~\eqref{eq:GPEOther} and \eqref{eq:GPECentral} in the mean-field limit.

The dark state exists for all parameter values, but its dynamic stability changes with $J$ and $\gamma$.
This stability determines the mean-field hysteresis illustrated in Fig.~\ref{fig:setup}(b), which compares the soliton bistability to conventional optical bistability. The hatched area indicates the difference in the area swept by the hysteresis loops. 
The lower branch of the soliton bistability features a sudden jump from zero (dark soliton) to a finite population at a threshold value $J=J_t$ (for a fixed $\gamma/\mu$), in keeping with a first-order phase transition.
The semi-classical transition, where quantum fluctuations are ignored, therefore corresponds to a first-order \textit{dark-state phase transition}, driven by a change in the stability of a dark state, as discussed in Refs.~\cite{Marcuzzi2015,Buchhold2017,Roscher2018,Sieberer2025}.\par

\begin{figure}[t]
    \hspace{-1em}
    \begin{minipage} {.33\linewidth}
    \vspace{-1em}
        \includegraphics[scale=1]{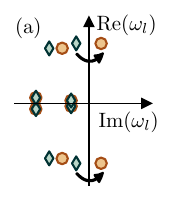}
    \end{minipage}
    \hspace{-1.15em}
    \begin{minipage} {.67\linewidth}
        \includegraphics[scale=1]{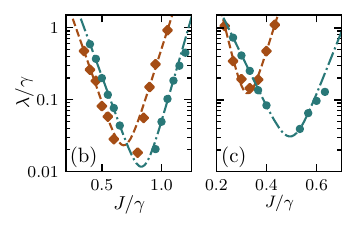}
    \end{minipage}
    \caption{(a) Linear-stability eigenvalues of the discrete dark soliton \eqref{eq:soliton} in the BH chain for $J/\gamma = 0.97$ (green) and $J/\gamma=1.06$ (orange). 
    (b),(c) Effective Liouvillian gap $\lambda$ as a function of $J/\gamma$ at fixed dissipation rates (b) $\gamma/\mu=0.1$ and (c) $\gamma/\mu=0.3$.
    Results with the soliton (green dots) are compared to those obtained with imposed mirror-symmetry (orange diamonds), which excludes the possibility of a soliton. Broken lines are fits to guide the eye. Simulations were performed with a fixed particle number per site and $L$ large enough to avoid boundary effects.}
    \label{fig:gapclosing}
\end{figure}

To probe the stability of the dark soliton we analyze the Bogoliubov spectrum in the $N\rightarrow\infty$ limit where the noise (dissipative and vacuum) vanishes. See Supplemental Material (SM) \cite{SM} for derivations. 
Figure \ref{fig:gapclosing}(a) shows the complex eigenvalues of the Bogoliubov modes for the two regimes. For $J < J_t$ (green), all eigenvalues $\omega_l$ have $\rm{Re}(\omega_l) < 0$, implying damped excitations and making the soliton \eqref{eq:soliton} a stable fixed point. For $J > J_t$ (orange), a pair of complex eigenvalues cross the real axis, indicating the onset of dynamical instability of the dark state. This signifies a \textit{Hopf bifurcation}, in contrast to the saddle-node bifurcations observed for optical bistability \cite{foss2017emergent}. The absence of a stable limit cycle indicates that the bifurcation is subcritical, leading the system to settle in the stable superfluid state.
Notably, the eigenvalues of these unstable modes have a non-zero real part, which implies 
an \textit{oscillatory} instability, as is typical of on-site solitons in 1D discrete lattices \cite{kivshar_dark_1994,johansson_discreteness-induced_1999}.

Corrections to the perfect dark state from quantum fluctuations vanish with $\mathcal{O}(1/N)$ (see SM \cite{SM} for details), leading to signatures of the dark state for large but finite systems. 
We will demonstrate its significance for experimentally relevant systems with hundreds of sites.
To capture the role of vacuum and dissipative noise in systems with finite atom numbers, we perform truncated Wigner approximation (TWA) simulations based on Eqs.~\eqref{eq:GPEOther} and \eqref{eq:GPECentral}. 
The inclusions of noise tends to smooth out the mean-field transition \cite{rodriguez2017probing,huybrechts_dynamical_2020}, but the extended hysteresis area remains.
Individual Wigner trajectories exhibit jumps between the dark soliton and the uniform superfluid, in analogy with optical bistability, triggered by the dynamic noise.
By extracting the characteristic switching time $\tau$ from large ensembles of trajectories, we estimate the Liouvillian gap $\lambda = 1/\tau$ \cite{Ceulemans2023}. Similar switching behavior was inferred experimentally for a comparable system in Ref.~\cite{benary_experimental_2022}.

Figures \ref{fig:gapclosing}(b) and (c) display the Liouvillian gap $\lambda$ as a function of the tunneling parameter $J$ at fixed dissipation rates (green). Results are also displayed with enforced mirror symmetry around the central site (orange) to prohibit soliton formation. 
The minimum gap, the best estimate for the critical point $J = J_c$, is seen to be lower for the soliton-superfluid transition, indicating 
fewer fluctuations.
The correspondingly slower switching rate is due to the fact that in order to switch from a soliton to the superfluid, the phase of the entire reservoir on the left or right 
must be flipped by $\pi$. 
In comparison, for optical bistability, switches in separate trajectories are driven by dynamic noise or fluctuations in the driving of a single site. The soliton 
bistability 
therefore requires increased fluctuations, as occurs at larger values of $J$.

This difference between the simulations with and without a soliton increases as the loss rate $\gamma$ grows, as shown by the minima in Fig.~\ref{fig:gapclosing}(c), due to the increased stability of the soliton. The curve also broadens significantly, 
reflecting a wider region around the minimum that is affected by critical slowing down. This is in contrast to  optical bistability, which sees a closing of this region with increasing $\gamma$ \cite{Drummond1980,Reeves2021}. 
Looking ahead, this is also shown in the bistability phase diagram, Fig.~\ref{fig:phasediagrams}(a)-(b), which demonstrates broadening of the bistable region in the $\gamma-J$ plane at larger loss rates only for the soliton bistability [Fig.~\ref{fig:phasediagrams}(a)].

\textit{Application to experiment.---}
We now demonstrate that the soliton dark-state phase transition quantitatively explains recent experimental observations in superfluid arrays with a single lossy site \cite{Labouvie2015,Labouvie2016,benary_experimental_2022,Rohrle2024}. These experiments were performed with ultracold rubidium in a quasi-1D Josephson chain, where each site still has substantial radial extent determined by a 2D isotropic harmonic trap in the $x$-$y$ plane with frequency $\omega_r$ [Fig.~\ref{fig:setup}(a)]. The discrete lattice is oriented in the $z$-direction, and the central site is subject to an electron beam \cite{Gericke2008,Wurtz2009} that introduces a controllable atom loss rate $\gamma$ \cite{Barontini2013}. 

\begin{figure}[t]
    \hspace{-1em}
    \begin{minipage}{0.465\linewidth}
        \includegraphics[width=\linewidth]{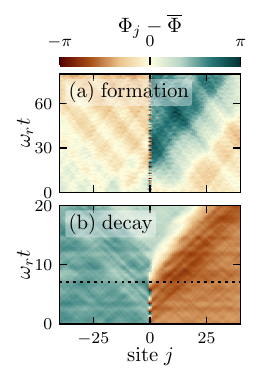}
    \end{minipage}
    \hspace{-1.3em}
    \begin{minipage}{0.585\linewidth}
        \vspace{1.5em}
        \includegraphics[width=\linewidth]{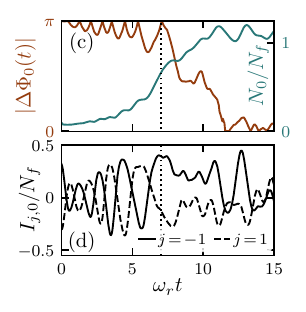}
    \end{minipage}
    \caption{Emergence and stability of a dark soliton in the experimental system. (a) Evolution of the spatial phase profile $\Phi_j$ vs time for a single trajectory with $L = 100$ (outer sites not shown), $|\Delta \Phi_0 (0)| \approx 0.7$, and all sites initially full, setting $\gamma/\mu=0.056$ and $J/\mu=0.007$.  A global phase $\bar{\Phi}$ is subtracted at each time for ease of visualization \cite{footnotephaseplot}. (b) As for (a) but starting from the soliton with initial $\Delta \Phi_0 (0) \approx \pi$ and low occupation on the lossy site, setting $\gamma = 0$ and $J/\mu = 0.08$.
    (c) Phase difference between driving sites $|\Delta \Phi_0 (t)|$ (left axis) and normalized atom number $N_0/N_f$ on the lossy site (right axis) vs time for the trajectory in (b).   
    (d) Corresponding time evolution of the current contributions from the respective driving sites, $I_{1,0}$ (dashed) and $I_{-1,0}$ (solid).}
    \label{fig:3D}
\end{figure}

These experiments typically operate with $\mathcal{O}(10^3)$ particles per site, such that we are able to approximate the dynamics using classical fields $\psi_j(\bm{x},t)$ evolving under coupled stochastic Gross-Pitaevskii equations \cite{blakie_projected_2005,Blakie2008,proukakis_quantum_2013}. In the TWA, time evolution of these fields is governed again by Eqs.~\eqref{eq:GPEOther} and \eqref{eq:GPECentral}, but with 
\begin{align}
\mathcal{H}\psi_j = \mathcal{P}\left\{\left[- \frac{\hbar \nabla^2}{2m} +V(\bm{x}) + g_2 |\psi_j|^2 \right]\psi_j \right\} ,
\label{eq:GPEoperator}
\end{align}
where $V(\bm{x}) = \frac{1}{2} m \omega_r^2 (x^2+y^2)$ and $g_2$ is the 2D interaction strength~\cite{SM}. The projection operator $\mathcal{P}$ imposes a cut-off that we use to restrict the numerics to modes at relevant energies; see Ref.~\cite{Blakie2008} and SM \cite{SM} for details. The complex Gaussian noise $\eta$ then satisfies $\langle \eta(\bm{x},t)\eta^*(\bm{x}^\prime,t^\prime)\rangle = \delta_C(\bm{x},\bm{x}^\prime)\delta(t-t^\prime)$,  where $\delta_C$ acts like the Dirac delta on the projected space \cite{Blakie2008}.
For a single experimental run, the system is initialized in its ground state, before atoms are removed from the central site (if required) using an electron beam.
During this initial state preparation the two halves of the system are disconnected and can drift out of phase. We approximate this effect by multiplying one half-chain by a random phase $e^{i\phi}$ \cite{SM}, a method used in Ref.~\cite{begg_nonequilibrium_2024} to obtain quantitative agreement with a comparable experiment \cite{Labouvie2015} focused on the dissipationless ($\gamma=0$) case.

We first demonstrate the emergence of the soliton for the experimental system. It has been argued that in the TWA formalism an individual trajectory is analogous to an individual experimental run \cite{Blakie2008}. Figure~\ref{fig:3D}(a) shows the evolution of the phase at each site $\Phi_j$ for an example single Wigner trajectory with strong dissipation and an initial state with all sites full and the phase difference between driving sites  $|\Delta \Phi_0| = | \Phi_1 - \Phi_{-1}|\ll \pi$ \cite{phasediff_footnote}. It can be seen that the system quickly relaxes to the soliton state with a $\pi$ phase difference from left to right.

Next, we demonstrate that despite the system's significant radial size, it is the 1D oscillatory instability of the dark soliton that sets the relevant timescale. Figure~\ref{fig:3D}(b) shows the evolution of phase $\Phi_j$ for a single trajectory starting from a dark soliton state but with no dissipation ($\gamma = 0$). The corresponding evolution of the phase difference $|\Delta\Phi_0|$ and the atom number on the lossy site, $N_0$, are displayed in Fig.~\ref{fig:3D}(c).  
The phase difference remains near $\pi$ until the time $t_r$ at which the occupation has grown significantly (dotted line). Around $t_r$, the location of the $\pi$ phase jump becomes unstable and propagates ballistically in a spontaneously chosen direction, seeded by quantum noise, leading to phase resynchronization.
The robustness of a soliton-like excitation up to time $t_r$ is consistent with the slow change in atom number compared to cases with smaller phase differences, as shown by Ref.~\cite{begg_nonequilibrium_2024}. The change in occupation over time is  
\begin{align}
\partial_t N_j  = 2J/\hbar \left( I_{j+1,j} + I_{j-1,j} \right), \label{eq:fillrate}
\end{align}
where $I_{k,j} \equiv \mathrm{Im}[\braket{\psi_{k}}{\psi_j}]$ denotes the current flowing between sites $j$ and $k$.
Figure~\ref{fig:3D}(d) shows the time evolution of the current contributions from the respective driving sites, $I_{-1,0}$ and $I_{1,0}$.
While the total current remains small for some time, indicated by the slowly changing atom number in Fig.~\ref{fig:3D}(c), the contributions from the left and right driving sites oscillate with approximately equal magnitude but opposite sign, characteristic of an oscillatory instability. This suggests that, despite each site being quasi-2D, it is the 1D physics that sets the dominant time-scale. The growth of these instabilities leads to filling of the site and consequent propagation of the phase slip in space (Fig.~\ref{fig:3D}(b)). This remains the case when dissipation is added, which further extends the soliton lifetime (see SM \cite{SM}).

\begin{figure}[t]
    \hspace{-1em}
    \includegraphics[scale=1]{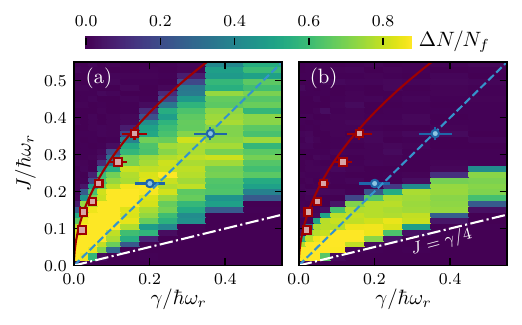}
    \caption{(a) Phase diagram in $J$-$\gamma$ parameter space, presented by the late time (normalized) atomic population difference $\Delta N$ between simulations starting from different initial states: the empty and full central site respectively. A finite $\Delta N$ indicates a bistable regime. The experimental phase boundaries from Ref.~\cite{Labouvie2016} are displayed with points and are fitted by $J \sim \sqrt{\gamma}$ (upper boundary) and $J \sim \gamma$ (lower boundary). The white dot-dash line indicates $J = \gamma/4$ as expected from theory. (b) As in (a) but for a system with imposed mirror symmetry around the central site, which is incapable of hosting a dark soliton. Displayed data corresponds to $t = 80$~ms and 251 lattice sites, averaged over 64 trajectories for each point.}\label{fig:phasediagrams}
\end{figure}

\textit{Phase diagram.---} We now construct the full bistability phase diagram by numerically solving Eqs.~\eqref{eq:GPEOther}, \eqref{eq:GPECentral}, and \eqref{eq:GPEoperator} for a system of 251 lattice sites using the TWA. The dynamics are evaluated at a fixed time $t = 80$ ms, starting from two distinct initial conditions: one with an initially empty central site and another with a full central site.
Figure~\ref{fig:phasediagrams}(a) shows the resulting atom number difference $|\Delta N|$ between these two initial conditions, as a function of the tunneling $J$ and dissipation $\gamma$. A large $|\Delta N|$ indicates sensitivity to initial conditions and thus signals bistability. For comparison, we overlay the experimental phase boundaries from Ref.~\cite{Labouvie2016}. Our simulations show excellent agreement with the experimental data at the upper boundary (large $J$), where the system transitions from a dark soliton to a uniform superfluid.
Notably, this transition has not been accurately reproduced in previous theoretical studies except by unrealistically adjusting experimental parameters \cite{Reeves2021,multimode_footnote}.

To highlight the role of the soliton, Fig.~\ref{fig:phasediagrams}(b) displays the results of simulations with imposed mirror symmetry. Enforcing perfect in-phase driving from both sides leads to a conventional optical bistability \cite{Labouvie2016,Reeves2021,Ceulemans2023}, which predominantly affects the upper boundary of the bistable region. These soliton free simulations predict an upper threshold at considerably \textit{lower} $J$ values than what is observed experimentally. 
While a lowering of the upper boundary can possibly be attributed to thermal and quantum fluctuations, it is much harder to account for an \textit{increase} of the upper phase boundary. 
We therefore conclude that the spatial phase structure of the soliton played a crucial role in the observed bistability. It would be interesting to confirm this prediction by directly measuring this phase difference in experiment.

Both phase diagrams differ significantly from the experimental data at the lower boundary. While theoretically this boundary is predicted to follow $J=\gamma/4$ \cite{Drummond1980,Reeves2021} (white dashed lines), a much steeper linear relation, $J=\gamma$, was found experimentally. 
We believe that this discrepancy can be explained by finite-size and temperature effects. In particular, the experimental system in Ref.~\cite{Labouvie2016} 
only extended across $\sim65$ full sites, albeit not uniformly occupied due to the weak harmonic confinement in the lattice direction. The sound velocity $v = \sqrt{2Jd_z^2\mu}$ implies that over these time-scales approximately 200 sites are needed to avoid finite-size effects. The dissipation could therefore have time to reduce the atom number of all sites substantially, lowering the interaction energy and making it easier to switch to the soliton state. Moreover, since an initial soliton state prevents substantial atom loss, the data at the upper boundary is far less affected by finite-size effects. In the SM \cite{SM} we confirm this by showing that the slope of the lower boundary increases with increasing fluctuations in the initial state, without substantially altering the agreement at the upper boundary.

\textit{Conclusion.---}
We have uncovered a semi-classical first-order dark-state phase transition between a dark soliton and uniform superfluid state, which we used to quantitatively explain a phase boundary in a recent cold atoms experiment \cite{Labouvie2016}. This result sheds new light on these and other experimental observations \cite{benary_experimental_2022,Rohrle2024}, previously not understood from first principles. We demonstrated this through comparison with the mirror symmetric system that displays a conventional optical bistability but is in stark disagreement with the experimental results. 
The transition can be understood via Bogoliubov theory and with the 1D Bose-Hubbard model, for which our results confirm that the region of critical slowing down is substantially broadened for the soliton bistability (as is seen in the phase diagram) and the critical point is shifted to higher tunnel couplings. 

Our results demonstrate that spatially extended coherent phase structures can fundamentally alter mean-field driven-dissipative transitions. 
Looking ahead, this mechanism suggests potential vortex–superfluid transitions in 2D, stabilization of phase structures via spatially patterned dissipation, and using dissipation as a phase switch for soliton imprinting \cite{Denschlag2000,Brazhnyi2009,Trimborn2011,Baals2021} or for controlling superfluid flow \cite{Will2023}. 
Additionally, there are potential links to two-terminal superfluid and superconducting junctions, where transport properties are substantially modified by single particle loss \cite{Ogino2025}.

\textit{Acknowledgements.---} 
We thank M. Reeves and H. Ott for many useful discussions. R.C. and M.W. acknowledge the support of the Research Foundation-Flanders (FWO) through Project No. 39532, the HPC core facility CalcUA and the Flemish Supercomputer Center (VSC). 
S.E.B. acknowledges the support of the Young Scientist Training Program at the Asia Pacific Center for Theoretical Physics, and that this work was performed with support from the National Science Foundation (NSF) through Award No. OMR-2228725. M.J.D. acknowledges funding from the Australian Research Council (ARC) Centre of Excellence in Future Low-Energy Electronics Technologies (Project No. CE170100039) and an ARC Discovery Project (Project No. DP250102923).

\bibliographystyle{apsrev4-2}


\begin{thebibliography}{70}
\bibitem [{foo()}]{footnotephaseplot}%
  \BibitemOpen
  \href@noop {} {}\bibinfo {note} {In Fig.~\ref{fig:3D}(a) we subtract the mean
  phase at each time, evaluated using the circular mean, for ease of
  visibility; this removes the rapidly oscillating global phase. However, for
  an initial soliton configuration as in Fig. ~\ref{fig:3D}(b), the circular
  mean vanishes because the two domains differ by approximately $\pi$. In this
  case, we instead subtract by the average phase of the end sites,
  $(\Phi_{-L/2} + \Phi_{L/2})/2$, which centers the soliton's phase in each
  domain on $\pm \pi/2$ respectively. We then correct for trivial global $\pi$
  shifts that can flip these phases, i.e. $\pm \pi/2 \rightarrow \mp \pi/2$,
  from one time-step to the next.}\BibitemShut {Stop}%
\bibitem [{pha()}]{phasediff_footnote}%
  \BibitemOpen
  \href@noop {} {}\bibinfo {note} {We define $\Delta \Phi_j = \Phi_{j+1} -
  \Phi_{j-1}$, where $\Phi_j = \langle \phi_{j+1}(\bm{x}) \rangle_{\bm{x}}$,
  and the spatial average of the local phase $\phi_{j}(\bm{x})$ is conducted
  over a central region of site $j$.}\BibitemShut {Stop}%
\end{thebibliography}

%

\end{document}


\title{Supplemental Material for ``Dark Soliton Formation as a Dark-State Phase Transition in a Dissipative Superfluid Josephson Junction Chain"}
\author{Robbe Ceulemans}
\affiliation{TQC, Universiteit Antwerpen, Universiteitsplein 1, 2610 Antwerpen, Beglium}
\author{Samuel E. Begg}
\affiliation{Department of Physics, Oklahoma State University, Stillwater, Oklahoma 74078, USA}
\affiliation{Asia Pacific Center for Theoretical Physics, Pohang 37673, Korea}
\affiliation{Department of Physics, The University of Texas at Dallas, Richardson, Texas 75080, USA}
\author{Matthew J. Davis}
\affiliation{Australian Research Council Centre of Excellence in Future Low-Energy Electronics Technologies, School of Mathematics and Physics, University of Queensland, St Lucia, Queensland 4072, Australia.}
\author{Michiel Wouters}
\affiliation{TQC, Universiteit Antwerpen, Universiteitsplein 1, 2610 Antwerpen, Beglium}
\email{Robbe.Ceulemans@uantwerpen.be}
\date{\today}

\maketitle

\section{Truncated Wigner approximation and semi-classical equations of motion}
We briefly discuss the derivation of 
Eqs. \ref{eq:GPEOther} and \ref{eq:GPECentral} 
in the main text via the Wigner phase space formalism. We consider the dissipative Bose-Hubbard model described by the Lindblad master equation
\begin{align}
\frac{d\hat{\rho}(t)}{dt} & =   -i [\hat{H}, \hat{\rho}(t)] +   \hat{\Gamma}\hat{\rho}(t) \hat{\Gamma}^{\dagger}   - \frac{1}{2} \{\hat{\Gamma}^{\dagger}\hat{\Gamma}, \hat{\rho}(t) \}, \label{eq:lindblad}
\end{align}
with Bose-Hubbard Hamiltonian 
\begin{align}
    \hat{H} = -J \sum_{j} \big(\hat{a}_{j+1}^{\dagger} \hat{a}_j + {\rm H.c.}\big) + \frac{U}{2}\sum_j \hat{a}^{\dagger}_j\hat{a}^{\dagger}_j \hat{a}_j  \hat{a}_j , \label{eq:bose_hubbard}
\end{align}
where $\hat{a}^{\dagger}_j$ and $\hat{a}_j$ respectively denote bosonic creation and annihilation operators for site $j \in \{ -L/2, L/2\}$ in a 1D chain of length $L+1$, and the particle loss on the central site is described by the jump operator $\hat{\Gamma} = \sqrt{\gamma} \hat{a}_0.$ The parameter $J$ is the nearest-neighbor tunneling amplitude and $U$ is the on-site interaction energy. 
Equation~\ref{eq:lindblad} can be represented in $2(L+1)$ dimensional phase space, where the Wigner function $W[\Psi,\Psi^*]$ is the analogue of the density matrix. The Wigner distribution evolves according to
\begin{equation}
\begin{split}
    \pdv{W}{t} &= -i\sum_j\left(\pdv{\psi_j}\mathcal{F}[\psi_j] - c.c.\right)W[\Psi,\Psi^*]\\
    &+ \frac{\gamma}{2}\pdv{}{\psi_0}{\psi_0^*}W[\Psi,\Psi^*] + \mathcal{O}(\partial^3),
\end{split}
\end{equation}
where
\begin{equation}
    \mathcal{F}[\psi_j] = J(\psi_{j-1}+\psi_{j+1}) - U(\abs{\psi_j}^2-1)\psi_j + i\frac{\gamma\delta_{0j}}{2}\psi_j.
\end{equation}
Neglecting third order derivative terms, which vanish with $\mathcal{O}(1/N)$, this reduces to a Fokker--Planck equation, which can be mapped to the Langevin 
Eqs. \ref{eq:GPEOther} and \ref{eq:GPECentral} 
of the main text. This also implies an equation for the evolution of the local density $n_j$, which in the limit $N \rightarrow \infty$ on every site is given by  
\begin{align}
  & i \frac{d}{dt}n_j   =  -J (\psi_{j-1} + \psi_{j+1}) \psi_j^*   + {\rm H.c.}    \label{eq:darkstate}   - i \delta_{0, j} \gamma_j n_j , 
\end{align}
where $\psi_j=\sqrt{n_j}e^{i\phi_j}$. The dissipative contribution vanishes for a dark soliton (Eq. \ref{eq:soliton} of the main text), since $n_0 = 0$. The conjugate equation for $\phi_0$ is irrelevant as this phase is purely a gauge due to the vanishing density. In addition, the dark soliton is an eigenmode of the remaining terms, i.e. the discrete Gross-Pitaevskii equation 
\begin{equation}
    i\hbar\partial_t\psi_j =  - J\left(\psi_{j-1} + \psi_{j+1}\right) + U\abs{\psi_j}^2\psi_j \label{eq:gross_pit}.
\end{equation}
The vanishing dissipation and eigenmode condition
make the dark soliton a dark state of the Lindblad master equation in this limit.

\section{Linear stability analysis}
\subsection{One dimension}
To determine the dynamical stability of the dark soliton in the dissipative Bose-Hubbard model we analyze the eigenvalues of Bogoliubov excitations on top of the stationary dark soliton. The relevant equations are obtained by substituting the ansatz $\psi_j = \phi_j + \delta\psi_j(t)$ in 
Eq.~\ref{eq:GPEOther} and \ref{eq:GPECentral} 
of the main text and expanding the small perturbation as
\begin{equation}
    \delta\psi_j(t) = \sum_l u_j^{(l)} e^{-i\omega_l t} + v_j^{(l)*} e^{i\omega_l^* t}.
\end{equation}
Here, $u_j^{(l)}$ and $v_j^{(l)}$ are the amplitudes of the $l$-th normal mode with associated mode frequency $\omega_l$. The resulting eigenvalue problem,
\begin{align}
        \omega_lu_{j,l} =&\big(-2J+2U\abs{\psi_j}^2-\mu-i\gamma/2\big)u_{j,l} \nonumber \\ &+ \psi_j^2v_{j,l} - J\big(u_{j-1,l} + u_{j+1,l}\big), \nonumber\\ 
        \omega_lv_{j,l} =& \big(2J - 2U\abs{\psi_j}^2 +\mu -i\gamma/2\big)v_{j,l}\nonumber \\
        & - \psi_j^2u_{j,l} + J\big(v_{j-1,l} + v_{j+1,l}\big), \label{eq:bog_eig}
\end{align}
is solved numerically, with $\phi_j$ given by Eq.~\ref{eq:soliton} in the main text. Two examples of the eigenvalue spectrum are depicted for values of $J/\gamma$ above and below the critical point in Fig.~\ref{fig:gapclosing} of the main text. These clearly indicate the dynamically (un)stable nature of the dark state through the positive imaginary part of the individual modes.

\begin{figure}[t]
    \centering
    \includegraphics[scale=1]{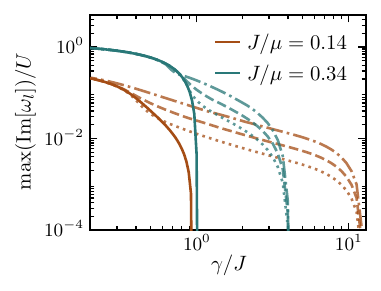}
    \caption{Imaginary value of the pair of unstable modes as a function of relative dissipation strength $\gamma/J$ (log-log scale). Different line styles indicate different system sizes; respectively $L = 100$ (solid), $L = 200$ (dash-dot) and $L=400$ (dashed) for the full, dash-dotted and dashed line. The dotted line illustrates the thermodynamic limit that was effectively realised with the incoherently pumped model introduced in Ref. \cite{Ceulemans2023}.}
    \label{Fig:CriticalLoss}
\end{figure}

Changes in the instability growth rate, $\mathrm{max}\left[\mathrm{Im}(\omega_l)\right]$, as a function of dissipation rate $\gamma$ are illustrated in Fig.~\ref{Fig:CriticalLoss}. The solid lines represent the predictions of the incoherent pumping model introduced in Ref.~\cite{Ceulemans2023} and show the swift decrease with an increasing dissipation rate. They clearly indicate, for a given tunneling strength $J$, a critical dissipation rate above which the dark soliton is rendered dynamically stable. These curves are complemented with results for finite Bose-Hubbard arrays (obtained via Eq. \ref{eq:bog_eig}) with hard-wall boundaries of different lengths $L$, denoted by the broken lines. For very small $\hbar \gamma/\mu$, the different curves overlap, but eventually branch off. 
With increasing $L$ the growth rate curves also overlap with the solid lines up until larger values of $\gamma/\mu$, indicating that the incoherent pumping model presented in \cite{Ceulemans2023} provides a plausible extrapolation to the thermodynamic limit. Furthermore, for a fixed dissipation rate in the intermediate regime of Fig.~\ref{Fig:CriticalLoss}, the soliton lifetime, given by the inverse of the instability growth rate, increases with growing system size $L$. 
Intuitively, for the finite-size systems, excitations created by the defect at the center of the chain can not leave the system and are able to destroy the soliton state after multiple reflections from the hard-wall boundaries. This effect is reduced for larger chains.
Notably, for large enough $\gamma/\mu$ the instability is still suppressed for the finite size systems.

\subsection{Three dimensions}
\begin{figure}[t]
    \centering
    \includegraphics[scale=1]{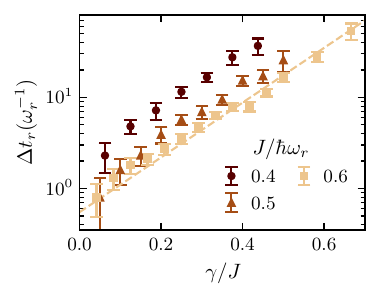}
    \caption{Shifted relaxation timescale $\Delta t_r =  t_r - t_r(\gamma=0)$ as a function of relative dissipation strength $\gamma/J$, with fit $\Delta t_r \sim 10^{6.28 \gamma/J}$.  Dots indicate mean result from 10 samples and error bars indicate standard error. Results correspond to different $J$ (legend), with $g_2 = 0.23 \hbar \omega_r$ and $L=100$. The state is initialized with $\Delta \Phi_0 (0) \approx \pi$ with each half of the system in its ground-state and an initial occupation on the central site of approximately $5\%$. }
    \label{fig:relaxation_time}
\end{figure}

We now discuss the stability for the case of the quasi-1D chain, where each site is 2D. In the main text, Fig.~\ref{fig:3D} showed that starting from an initial state close to the soliton configuration, an oscillatory instability eventually leads to the erasure of the soliton, just as in the 1D case discussed above. These simulations were carried out in the absence of dissipation. We now show that dissipation increases the stability of the soliton. We define the corresponding relaxation time $t_r$ as the first time in which the density on the central site exceed two-thirds of the value of a full site $N_f$, which indicates a significant departure from the soliton state in its evolution towards a `full' central site. This aligns with the definition of `filling time' in recent work by some of us \cite{begg_nonequilibrium_2024} on unitary dynamics in a similar set-up. 
Figure~\ref{fig:relaxation_time} shows $t_r$ as a function of $\gamma$. It can be seen that the relaxation time diverges exponentially with $\gamma$ as $\omega_r \Delta t_r = \omega_r [t_r - t_r(\gamma=0)] \sim 10^{6.28\gamma/J}$, indicating that for fixed $J$ the soliton becomes increasingly stable at  higher dissipation rates due to the suppression of the oscillatory instability.

\section{Optical bistability}
In the main text, we mention that the mirror-symmetric system possesses a conventional optical bistability. This follows from observations of Ref. \cite{Reeves2021}, which developed a mean-field theory for the central site driven by a single undepletable coherent reservoir $\psi_R$ and subject to dissipation. This is analogous to the situation in optical bistability of driven-dissipative Kerr resonators \cite{Drummond1980b}. The equations of motion for the field $\psi_0$ are 
\begin{align}
    i\hbar \frac{d\psi_0}{dt} = \mu \psi_0 + U |\psi_0|^2 \psi_0 - 2 J \psi_R - i \frac{\gamma}{2}\psi_0, 
\end{align}
where $\mu$ is the chemical potential on the central site. The steady-state solution possesses regions with one or three solutions; the latter being the bistable region. A pitchfork bifurcation occurs at the transition from one to multiple solutions. See Ref. \cite{Reeves2021} for additional details. 

\section{Details of experiment and simulations}
\subsection{Experiment} \label{sec:param}
As discussed in the main text, the experiment of Ref. \cite{Labouvie2016} involved a one-dimensional chain of two-dimensional harmonic traps. 
Following an initial cooling process that produces the cigar-shaped bulk condensate, a one-dimensional optical lattice is applied with a depth of $s=30~E_r$ (where $E_r$ is the lattice recoil energy), effectively isolating every site in a deep well for approximately 9~ms. During this preparation stage, the density on the central site can be independently adjusted relative to the rest of the system. 
To demonstrate bistability, two different initial states are distinguished. The local dissipation mechanism is either activated or left off during the preparation stage, leading to two possible scenarios: the trap at the center of the chain is either almost completely depleted or retains an occupation similar to the surrounding sites. 
Subsequently, the lattice is ramped down to a value within the range $s=6-12~E_r$, initiating the system dynamics. At this lower lattice depth the electron beam is turned on to provide a continuous source of particle loss.

The lattice spacing in the $z$-direction is $a = 549$~nm and the radial trap-frequency is $\omega_r= 165 \times 2\pi $ rad/s.  The total number of Rb-87 atoms is $45 \times 10^3$, with mass $m=1.44 \times 10^{-25}~{\rm kg}$ and an s-wave scattering length $a_s = 5.8 $ nm. Traps near the center of the chain contain approximately $N = 700$ atoms. The setup is harmonically trapped in the $z$-direction as well, with $\omega_z \approx 10 \times 2\pi~\mathrm{rad/s} \ll \omega_r$.

\subsection{Simulation parameters and procedure} \label{sec:param}
For our simulations we neglect the slowly varying harmonic trap along the $z$-axis and simulate a uniform lattice, as in our recent work in Ref.~\cite{begg_nonequilibrium_2024}. We begin with a first principles calculation of the corresponding lattice potential in the $z$-direction, $V(z)=V_0 \sin^2(k z)$, where $k= 2 \pi/\lambda$ is the lattice wavevector with wavelength $\lambda = 2a$ and spacing $a$. In the lowest band approximation the on-site potential can be approximated as a harmonic oscillator, $V(z) \approx \hbar^2 \omega_{\mathrm{lat}}^2z^2/4 E_r$, where $E_r = \hbar^2 k^2/2m$ is the lattice recoil energy.
Using the tight-binding approximation, the tunneling can be obtained as $J = \Delta E/4$ where $\Delta E$ is the bandwidth of the lowest Bloch band \cite{Arzamasovs2017} which we obtain by diagonalizing $\hat{H} = - \hslash^2 \partial_z^2/2m + V_0 (2 \pi z / \lambda)^2$ numerically. The tunnel coupling $J$ and $\omega_\mathrm{lat}$ are in one-to-one correspondence and we restrict our attention to the experimentally relevant ranges \cite{Labouvie2016}.

For a given $\omega_\mathrm{lat}$ the interaction strength and chemical potential can be computed as follows. Firstly, the 3D interaction strength is given by $g = 4\pi\hbar^2 a_s/m$. Due to the tight confinement in the $z$-direction, we model this direction as discrete. The effective 2D interaction strength for each trap is given by numerically integrating over the Wannier function $w(z)$ in the $z$-direction, i.e. $g_2 = g \int dz ~ w(z)^4 .$ The chemical potential is then estimated as $\mu =\sqrt{g_2 m \omega_r^2 N /\pi}$ using the Thomas-Fermi approximation. This procedure uniquely determines $\omega_\mathrm{lat}$, $g_2$, and $\mu$. 

To find the ground state we evolve the full chain in imaginary time. For the case in which the central site is initially full, this is taken as the initial condition. In contrast, for the complementary initial state we decrease the atom number on the central site to  $5\%$ of its full value. 
As discussed in our recent work \cite{begg_nonequilibrium_2024}, the experimental preparation of the initially empty site introduces additional dephasing into the system. Specifically, both halves of the system are isolated from each other while the atom number is reduced on the central site. Ref.~\cite{begg_nonequilibrium_2024} demonstrated numerically that the preparation stage is sufficiently long to allow significant diffusion of the relative phases. Moreover, once the lattice depth is lowered again, the empty site ensures that they cannot immediately `reconnect', thereby allowing additional time for diffusion. 
To capture this effect, for  each Wigner trajectory we multiply the right-hand subset of the chain, $R= \{j|j>0\}$, by random phase $e^{i\theta}$ with $\theta \in [0,2\pi]$. This phase randomization was shown by Ref. \cite{begg_nonequilibrium_2024} to be necessary to achieve quantitative agreement with the experimental results of a comparable setup, albeit one governed by unitary dynamics. For the dissipative system the effect of this random phase is less significant. 
For the case of an initially full-site configuration, the choice of initial phase difference $\theta$ has no effect on the long-time results.

\subsection{Simulation details}
In this section we provide a brief summary of the classical field (c-field) numerical methods employed in the main text. For the complete three-dimensional configuration, where each trap is modeled as a two-dimensional system, we need to restrict the number of modes per trap to satisfy the c-field approximation. This truncation is performed via the projection of the classical fields onto the coherent region $\mathcal{C}$, defined by 
\begin{align}
\mathcal{P}\{f(\bm{x})\} = \sum_{\bm{n} \in \mathcal{C}} \varphi_{\bm{n}} (\bm{x}) \int d^2 \bm{x}' \varphi_{\bm{n}}^*(\bm{x}') f(\bm{x}').
\label{eq:Projector}
\end{align}
Here, $\varphi_{\bm{n}}$ are eigenmodes of the cartesian Hermite-Gauss computational basis with quantum numbers $\bm{n} \in \{n_x,n_y\}$, which diagonalize the non-interacting problem ($g_2 = 0$). 
Letting $n = n_x + n_y$, we choose a cut-off $n_{\rm cut}$ to define the ``classical field region" $\mathcal{C} = \{n \; | \; n_x + n_y \leq n_{\rm cut}\}$. 
This approach incorporates a consistent energy cut-off that limits the number of radial modes for each trap.
In practice, we use a cut-off of approximately $n_{\rm{cut}} = 2.5\mu/\hbar \omega_r$.
Previous works \cite{Reeves2021,begg_nonequilibrium_2024} have shown that results for this system are largely independent of the exact cutoff for $n_{\rm cut} > 2\mu/\hbar\omega_r$. This cut-off captures scattering processes in which two atoms at the chemical potential energy collide, leading to one atom with energy $2\mu$ and another with approximately zero energy. This is anticipated to be relevant when the lossy site is significantly depleted, with an energy that is lower than neighboring sites by approximately $\mu$. Atoms therefore cannot tunnel directly into the depleted condensate, but can undergo tunneling into excited states of the trap and relax via collisions.

Having defined a classical field region, the coupled projected Gross-Pitaevskii equations, displayed in Eqs.~\eqref{eq:GPEOther} and \eqref{eq:GPECentral} 
in combination with 
Eq.~\eqref{eq:GPEoperator} 
of the main text, are integrated in the Hermite-Gauss basis using Gaussian quadrature~\cite{Blakie2008,Boyd2001}. 
For the simulations presented in the main text, the initial state is assumed to be a multi-mode coherent state $\psi_j(\bm{x},0) = \sum_{\bm{n}\in\mathcal{C}}\alpha_{j,\bm{n}}\varphi_{\bm{n}}(\bm{x})$, with the coefficients $\alpha_{\bm{n}}$ sampled from the phase space distribution
\begin{equation}
    W_c[\psi_j,\psi_j^*] = \left(\frac{2}{\pi}\right)^{\abs{\mathcal{C}}}\prod_{\bm{n}\in\mathcal{C}}\exp\left(-2\abs{\alpha_{j,\bm{n}} - \alpha^{(0)}_{j,\bm{n}}}^2\right),\label{eq:initialwigner}
\end{equation}
where $|\mathcal{C}|$ is the number of modes in the coherent region.
Here, $\alpha^{(0)}_{j,\bm{n}}$ are the mean-field values obtained through imaginary time evolution for a given chemical potential.
For a broader overview of c-field methods see Ref. \cite{Blakie2008}. 

The numerical execution of the method is verified using two independent implementations, one coded in Julia \cite{bezanson_julia_2017} by the author R.C., and another using the XMDS \cite{xmds2} software. Both implementations yield identical results.

\section{Finite-size and temperature effects}
\begin{figure}[t]
    \centering
    \includegraphics[scale=1]{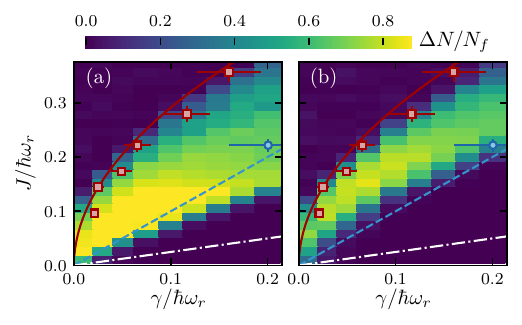}
    \caption{Phase diagram in $J$-$\gamma$ parameter space for $L = 70$, with increased uncertainty on initial state sampling as compared to the simulations in the main text: the standard deviation of the initial noise distribution is multiplied by (a)~$1.5$ and (b)~$2$. The color indicates the late time (normalized) atomic population difference $\Delta N$ (legend) between simulations starting from different initial states: the empty and full central site respectively. Displayed data corresponds to $t = 80$~ms, averaged over 64 trajectories for each point. For comparison, the experimental phase boundaries from Ref.~\cite{Labouvie2016} are displayed with points and are fitted by $J \sim \sqrt{\gamma}$ (upper boundary) and $J \sim \gamma$ (lower boundary). The white dot-dash line indicates $J = \gamma/4$ as expected from theory.}
    \label{fig:fluctuations}
\end{figure}

In this section we demonstrate that the discrepancy with experimental results at the lower boundary of the phase diagram, shown in Fig. \ref{fig:phasediagrams}(a) of the main text, could potentially be accounted for by additional fluctuations specific to the experimental setup. To mimic the effect of fluctuations due to a finite temperature, we increased the variance on the distribution for the initial conditions in Eq.~\eqref{eq:initialwigner}. Formally, this does not correspond to a finite temperature, but still provides a qualitative picture of how the phase boundaries would change. 

Fig. \ref{fig:fluctuations} shows the bistability phase diagrams for the cases where the standard deviation of the initial distribution is multiplied by (a) $1.5$ and (b) $2$. Additionally, a smaller system with only 70 sites is simulated, in line with the experiment \cite{Labouvie2016}. Notably, the largest change can be observed in the slope of the lower boundary. Enhanced fluctuations raise the lower boundary, but have no significant effect on the upper boundary, in agreement with observations in Ref. \cite{Reeves2021}. 
We conclude that the smaller system size in combination with fluctuations specific to the experimental setup cause the slope of the lower boundary to increase from the theoretical prediction of $J=\gamma/4$ \cite{Drummond1980b,Reeves2021} (white dashed-dotted line) to approximate the observed $J=\gamma$ (blue dashed line). The upper boundary is less affected due to the robust nature of the dark soliton. 

In addition, as discussed in the main text, the lower-boundary is expected to be far more sensitive to finite-size effects at a given time, since an initial soliton state prevents substantial atom loss as compared to the initially uniform superfluid state.

\bibliographystyle{apsrev4-2}


%